\documentclass[12pt]{article}
\usepackage{psfrag}
\usepackage{graphicx}
\usepackage{latexsym,amsfonts}
\usepackage{amsmath}
\usepackage{amssymb}
\usepackage{graphicx}
\usepackage{graphics}
\begin{document}
\setcounter{page}{1}
\def\theequation{\arabic{section}.\arabic{equation}}
\def\theequation{\thesection.\arabic{equation}}
\setcounter{section}{0}

\title{Registration of  hydrogen--like leptonic bound states ($e^-\mu^+$) and
($e^+\mu^-$) in reactions of high--energy scattering of polarized
electrons and positrons by nuclei with $Z \sim 100$ and analysis of CPT
invariance}

\author{E. A. Choban ~and~ V. A. Ivanova\,\thanks{E--mail:
 viola@kph.tuwien.ac.at}}

\date{\today}

\maketitle

\begin{center}
{\it Department of Nuclear Physics, State Polytechnical University of
St. Petersburg,\\ Polytechnicheskaya 29, 195251 St. Petersburg,
Russian Federation}
\end{center}

\begin{center}
\begin{abstract}
The cross sections for the reactions of muonium(anti--muonium)
production in high energy electron(positron) scattering by nuclei
$e^-(e^+) + Z \to Z + M^0(\bar{M}^0) + \mu^-(\mu^+)$ are calculated in
dependence on an energy and polarization of an initial
electron(positron) and a polarization of a final
$\mu^-(\mu^+)$--meson. Due to coherent phenomenon the cross sections
are proportional to $Z^2$. For $Z \sim 100$ due to the factor $Z^2$
the cross sections are large enough to be measured at energies
available for the HERA Collider at DESY. The results are discussed in
connection with a test of CPT invariance.
\end{abstract}
\end{center}

PACS: 11.10.--z, 13.40.--f, 13.66.De, 13.66.Jn

\newpage

\section{Introduction}
\setcounter{equation}{0}

\hspace{0.2in} The Standard Model \cite{KH02} represents the
Lagrangian approach \cite{SW95} to the description of strong,
electromagnetic and weak interaction of elementary particles, based on
the assumptions of locality and Lorentz invariance. Due to the
L\"uders--Pauli theorem (or the CPT theorem) \cite{AW80} locality and
Lorentz invariance of the Lagrangian of a quantum system lead to the
invariance of a quantum system under CPT transformation which contains
(i) a charge conjugation (C), a replacement of all particles by their
anti--particles, (ii) a parity transformation (P), a reflection of
spatial coordinates $(t, \vec{x}\,) \to (t, -\vec{x}\,)$, and (iii) a
time reversal (T), a reflection of time $(t, \vec{x}\,) \to (- t,
\vec{x}\,)$. A simplest consequence of the CPT theorem is the equality
of masses and lifetimes of particles and their anti--particles. At
present these are the most well verified experimentally requirements
of the CPT theorem \cite{PDG1}. Nevertheless, theoretical and
experimental test for CPT invariance is still a well motivated problem
of Elementary particle and Nuclear Physics \cite{LS69}. This is
related to the development of modern quantum field theories of strings
and superstrings \cite{MG86}, which are more fundamental than the
Standard Model and include it in the low--energy limit. Since string
theories deal with extended non--local objects, the L\"uders--Pauli
theorem is not valid for these theories. A direct consequence of this
can be a violation of CPT invariance for high energy reactions of
elementary particles and nuclei.

The problem of a test of CPT and Lorentz invariance has been recently
discussed by Kosteleck$\acute{\rm y}$ with co--workers
\cite{VK91}. They suggested to check CPT and Lorentz invariance
analysing a microwave spectroscopy of muonium $M^0$
\cite{GL77,VH92}. Muonium $M^0$ is a leptonic hydrogenlike bound state
of a positively charged muon $\mu^+$ and electron $e^-$.  It was
discovered in 1960 through the observation of its characteristic
Larmor precession in a magnetic field \cite{VH92}. The mean lifetime
of muonium $\tau_{M^0}$ is approximately equal to the lifetime of a
positively charged muon $\tau_{M^0} \simeq \tau_{\mu^+} = 2.197\times
10^{-6}\,{\rm s}$ \cite{KH02}. Due to absence of strong interactions
muonium is an ideal system (i) for determining of the properties of
muons, (ii) for testing of quantum electrodynamics \cite{CPT1}, and
(iii) for searching for effects of unknown interactions in the
electron--muon bound state \cite{CPT2}. Anti--muonium $\bar{M}^0$ is a
leptonic analog of anti--hydrogen. It is a bound state of a negatively
charged muon $\mu^-$ and positron $e^+$.

A hydrogenlike structure of muonium allows to use atomic notations for
the classification of its quantum states. For example, ${^{2S +
1}}L_J$ corresponds to the quantum state of muonium (or
anti--muonium) with a total angular momentum (or a total spin) $J$,
an angular momentum $L$ and a spin $S$ \cite{KH02}.

The use of muoniums $M^0$ and anti--muoniums $\bar{M}^0$\,
\footnote{Anti--muonium $\bar{M}^0$ is a bound state of a negatively
charged muon $\mu^-$ and a positron $e^+$. It is a leptonic analogy of
the anti--hydrogen.} as a laboratory for a test of CPT invariance has
been recently suggested by Choban and Kazakov \cite{EC01}. In their
approach muoniums and anti--muoniums are produced with a total angular
momentum $J = 0$ in the reactions $e^- + Z \to Z + M^0 + \mu^-$ and
$e^+ + Z \to Z + \bar{M}^0 + \mu^+$ of high energy scattering of
electrons and positrons by nuclei with a number of protons
$Z$. According to atomic classification muonium (or anti--muonium)
with a total angular momentum $J = 0$ can be in two bound states: (i)
a ground $1s$ state ${^1}{\rm S}_0$ with $L = S = 0$ and (ii) an
excited $2p$ state ${^3}{\rm P}_0$ with $L = S = 1$.

Due to principle of superposition muonium and anti--muonium should be
produced in the reactions $e^- + Z \to Z + M^0 + \mu^-$ and $e^+ + Z
\to Z + \bar{M}^0 + \mu^+$ in both states ${^1}{\rm S}_0$ and
${^3}{\rm P}_0$. The interference of these states should lead to
time--oscillations of a probability of muonium (anti--muonium)
detected at a moment $t$. A comparison of time--oscillations of
probabilities of the detected muonium and anti--muonium should testify
whether CPT invariance conserved or not. This is Choban--Kazakov's
idea of a test of CPT invariance in the high--energy reactions $e^- +
Z \to Z + M^0 + \mu^-$ and $e^+ + Z \to Z + \bar{M}^0 + \mu^+$. In
terms of formulas it can be represented as follows.

Let the wave function of muonium produced in the reaction $e^- + Z \to
Z + M^0 + \mu^-$ be defined by
\begin{eqnarray}\label{label1.1}
\Psi_{M^0}(t,\vec{x}\,) =\sqrt{\frac{m_{M^0}}{|\vec{k}\,|}}\,
e^{\textstyle i\,\vec{k}\cdot \vec{x} - i\,Et}\,\Psi_{M^0}(t),
\end{eqnarray}
where $E$ and $\vec{k}$ are an energy and 3--momentum of muonium,
$m_{M^0}$ is a mass of muonium. Note, that the energy $E$ does not
contain the contributions of the binding energies $E_{1s}$ and
$E_{2p}$ of the bound $1s$ and $2p$ states. The wave function
$\Psi_{M^0}(t)$ can be described by
\begin{eqnarray}\label{label1.2}
\Psi_{M^0}(t) = C_{1s}\,\exp\Big( -i\,\frac{m_{M^0}}{E}\,E_{1s}t\Big)
+ C_{2p}\,\exp\Big( -i\,\frac{m_{M^0}}{E}\,E_{2p}t\Big).
\end{eqnarray}
The coefficients $C_{1s}$ and $C_{2p}$ describe the contributions of
the $1s$ and $2p$ states, respectively. 

Introducing a parameter $\varepsilon = |C_{2p}|^2/|C_{1s} +
C_{2p}|^2$, related to a fraction of the excited $2p$ state in the
wave function of muonium $\Psi_{M^0}(t,\vec{x}\,)$ \cite{EC01}, the
probability to find muonium at the moment $t$ can be given by
\begin{eqnarray}\label{label1.3}
P_{M^0}(t) = P_{M^0}(0)\,[1 - 4\sqrt{\varepsilon}\,(1 -
\sqrt{\varepsilon})\sin^2(\Omega t)],
\end{eqnarray}
where $\Omega = m_{\mu}(E_{2p} - E_{1s})/2E = 5.103\times
10^{-6}\,(m_{\mu}/E)\,{\rm MeV}$ \cite{VK91}\,\footnote{The account
for a constant relative phase $2\varphi$ of coefficients $C_{1s}$ and
$C_{2p}$ changes the probability (\ref{label1.3}) as follows
$P_{M^0}(t) = P_{M^0}(0)\,[1 - 4\sqrt{\varepsilon}\,(\sqrt{1 -
\varepsilon\,\sin^2\varphi}-\sqrt{\varepsilon}\,\cos\varphi)\,
\sin(\Omega t + \varphi)\sin(\Omega t)]$}.

It is seen that the probability $P_{M^0}(t)$ is an
oscillating function.  A period of oscillations $T_{M^0}$ is
determined by
\begin{eqnarray}\label{label1.4}
T_{M^0} = \frac{2\pi}{\Omega} =\frac{4\pi}{E_{2p} -
E_{1s}}\,\Big(\frac{E}{m_{\mu}}\Big) = 1.232\times
10^6\,\Big(\frac{E}{m_{\mu}}\Big)\,{\rm MeV^{-1}}.
\end{eqnarray}
In order to get $T_{M^0}$ in seconds we have to multiply the r.h.s. of
(\ref{label1.4}) by $\hbar = 6.582\times 10^{-22}\,{\rm MeV\,s}$
\cite{VK91}. This yields $T_{M^0}= 8.106\times
10^{-16}\,(E/m_{\mu})\,{\rm s}$.  The period of oscillations $T_{M^0}$
should be compared with the lifetime of muonium in the laboratory
frame $t_{M^0}$ which is related to the mean lifetime $\tau_{M^0}$ by
the relativistic relation
\begin{eqnarray}\label{label1.5}
t_{M^0} = \Big(\frac{E}{m_{\mu}}\Big)\,\tau_{M^0}.
\end{eqnarray}
Taking into account that $\tau_{M^0} \simeq 2.197\times 10^{-6}\,{\rm
s} $ we are able to estimate the number of oscillations $\nu_{M^0}$:
\begin{eqnarray}\label{label1.6}
\nu_{M^0} = \frac{t_{M^0}}{T_{M^0}} \simeq 2.710\times 10^{\,9}.
\end{eqnarray}
The analogous expression can be written down for the probability
$P_{\bar{M}^0}(t)$ to detect anti--muonium at moment t with parameters
$\bar{\varepsilon}$ and $\bar{\Omega}$. The result reads
\begin{eqnarray}\label{label1.7}
P_{\bar{M}^0}(t) = P_{\bar{M}^0}(0)\,[1 -
4\sqrt{\bar{\varepsilon}}\,(1 -
\sqrt{\bar{\varepsilon}})\sin^2(\bar{\Omega} t)].
\end{eqnarray}
A relation of the probabilities $P_{M^0}(t)$ and $P_{\bar{M}^0}(t)$ to
the experimental analysis of the violation of CPT invariance in the
reactions $e^- + Z \to Z + M^0 + \mu^-$ and $e^+ + Z \to Z +
\bar{M}^0 + \mu^+$ is the following.

For the calculation of the amplitude of muonium and anti--muonium
production in the reactions $e^- + Z \to Z + M^0 + \mu^-$ and $e^+ + Z
\to Z + \bar{M}^0 + \mu^+$ we use the effective Lagrangian of the
$M^0\mu^+e^-$ interaction which can be defined as
\begin{eqnarray}\label{label1.8}
{\cal L}_{M^0\mu^+ e^-}(x) = g_{1s}\,\bar{\psi}_{\mu^-}(x)\gamma^5
\psi_{e^-}(x)\,\Phi^{\dagger}_{1s}(x) +
g_{2p}\,\bar{\psi}_{\mu^-}(x)\psi_{e^-}(x)\,\Phi^{\dagger}_{2p}(x),
\end{eqnarray}
where $\bar{\psi}_{\mu^-}(x) $ and $\psi_{e^-}(x)$ are local
interpolating fields of the $\mu^+$--meson and electron $e^-$,
$\Phi_{1s}(x)$ and $\Phi_{2p}(x)$ are local operators of interpolating
fields of muonium in the states $1s$ and $2p$, respectively. They are
expanded into plane waves and operators of creation and
annihilation. 

The wave functions of the relative motion of the muon $\mu^+$ and the
electron $e^-$ contribute to the coupling constants $g_{1s}$ and
$g_{2p}$, which define the interaction of muonium in the $1s$ and $2p$
states with $\mu^+e^-$ pair, respectively.  For the calculation of the
effective coupling constant we would use the wave functions of muonium
in the states ${^1}{\rm S}_0$ and ${^3}{\rm P}_0$ with a total
momentum $\vec{P}$ defined by \cite{SS61,IZ80}
\begin{eqnarray}\label{label1.9}
&&|M^0(\vec{P}\,);{^1}{\rm S}_0\rangle = \frac{1}{(2\pi)^3}\int
 \frac{d^3k}{\sqrt{2 E_{e^-}(\vec{k})}}\frac{d^3q}{\sqrt{2
 E_{\mu^+}(\vec{q})}}\,\delta^{(3)}(\vec{P} - \vec{k} -
 \vec{q})\,\varphi_{1s}(\vec{k})\nonumber\\
 &&\times\,\frac{1}{\sqrt{2}}\, [b^{\dagger}_{e^-}(\vec{k},+1/2)
 d^{\dagger}_{\mu^+}(\vec{q},-1/2) - b^{\dagger}_{e^-}(\vec{k},-1/2)
 d^{\dagger}_{\mu^+}(\vec{q},+1/2)]|0\rangle,\nonumber\\
 &&|M^0(\vec{P}\,);{^3}{\rm P}_0\rangle = \frac{1}{(2\pi)^3}\int
 \frac{d^3k}{\sqrt{2 E_{e^-}(\vec{k})}}\frac{d^3q}{\sqrt{2
 E_{\mu^+}(\vec{q})}}\,\delta^{(3)}(\vec{P} - \vec{k} -
 \vec{q})\,\varphi_{2p}(\vec{k})\nonumber\\ &&\times\,
 \frac{1}{\sqrt{2}}\,[b^{\dagger}_{e^-}(\vec{k},+1/2)
 d^{\dagger}_{\mu^+}(\vec{q},-1/2) + b^{\dagger}_{e^-}(\vec{k},-1/2)
 d^{\dagger}_{\mu^+}(\vec{q},+1/2)]|0\rangle,
\end{eqnarray}
where $|0\rangle$ is the vacuum wave function;
$b^{\dagger}_{e^-}(\vec{k},\sigma)\,(b_{e^-}(\vec{k},\sigma)$ and
$d^{\dagger}_{\mu^+}(\vec{k},\sigma)\,(d_{\mu^+}(\vec{k},\sigma)$ are
operators of creation (annihilation) of electron and muon $\mu^+$ with
a momentum $\vec{k}$ and polarization $\sigma = \pm 1/2$. These
operators obey the covariant canonical anti--commutation relations
\begin{eqnarray}\label{label1.10}
\{b_{e^-}(\vec{k},\sigma),b^{\dagger}_{e^-}(\vec{k}\,',\sigma\,'\,)
=(2\pi)^3 2E_{e^-}(\vec{k})\,\delta^{(3)}(\vec{k} -
\vec{k}\,')\delta_{\sigma \sigma\,'},\nonumber\\
\{d_{\mu^+}(\vec{k},\sigma),d^{\dagger}_{\mu^+}(\vec{k}\,',\sigma\,'\,)
=(2\pi)^3 2E_{\mu^+}(\vec{k})\,\delta^{(3)}(\vec{k} -
\vec{k}\,')\delta_{\sigma \sigma\,'}.
\end{eqnarray}
Then, $\varphi_{1s}(\vec{k})$ and $\varphi_{2p}(\vec{k})$ are the wave
functions of the $1s$ and $2p$ states in the momentum
representation. They are normalized to unity
\begin{eqnarray}\label{label1.11}
\int \frac{d^3k}{(2\pi)^3}|\varphi_{1s}(\vec{k})|^2 = \int
\frac{d^3k}{(2\pi)^3}|\varphi_{2p}(\vec{k})|^2 = 1.
\end{eqnarray}
The wave functions (\ref{label1.9}) are normalized by
\begin{eqnarray}\label{label1.12}
\langle {^1}{\rm S}_0; M^0(\vec{P}\,)|M^0(\vec{P}\,');{^1}{\rm
S}_0\rangle &=& (2\pi)^3
2E^{(1s)}_{M^0}(\vec{P})\,\delta^{(3)}(\vec{P} -
\vec{P}\,'),\nonumber\\ \langle {^3}{\rm P}_0;
M^0(\vec{P}\,)|M^0(\vec{P}\,');{^3}{\rm P}_0\rangle &=& (2\pi)^3
2E^{(2p)}_{M^0}(\vec{P})\,\delta^{(3)}(\vec{P} - \vec{P}\,'),
\end{eqnarray}
where $E^{(n)}_{M^0}(\vec{P}) = \sqrt{(m_{\mu^+} + m_{e^-} + E_n)^2 +
\vec{P}^{\;2}}$ is the total energy of muonium with $E_n = E_{1s}$ and
$E_n = E_{2p}$ for the $1s$ and $2p$ state, respectively.

In the limit $m_e \to 0$ due to invariance of the interpolating
electron field $\psi_{e^-}(x)$ under $\gamma^5$--transformation,
$\psi_{e^-}(x) \to \gamma^5 \psi_{e^-}(x)$, the effective Lagrangian
(\ref{label1.8}) can be transcribed into the form
\begin{eqnarray}\label{label1.13}
{\cal L}_{M^0\mu^+ e^-}(x) =\bar{\psi}_{\mu^-}(x)\gamma^5
\psi_{e^-}(x)\,(g_{1s}\,\Phi^{\dagger}_{1s}(x) +
g_{2p}\,\Phi^{\dagger}_{2p}(x)).
\end{eqnarray}
Through the loop diagrams in Fig.1 the coupling constants $g_{1s}$ and
$g_{2p}$ are related to the constants $C_{1s}$ and $C_{2p}$
(\ref{label1.2})

Since one cannot distinguish experimentally the $1s$ and $2p$ states
of muonium and of anti--muonium, the number of favourable events
$N_{M^0}(T)$ and $N_{\bar{M}^0}(T)$, detected during an interim $T$,
should be proportional to $\sigma^{(e^-\,Z)}_{M^0}(E_1)P_{M^0}(T)$ and
$\sigma^{(e^+\,Z)}_{\bar{M^0}}(E_1)P_{\bar{M}^0}(T)$
\begin{eqnarray}\label{label1.14}
N_{M^0}(T) &=& \sigma^{(e^-\,Z)}_{M^0}(E_1)P_{M^0}(T)LT,\nonumber\\
N_{\bar{M}^0}(T) &=&
\sigma^{(e^+\,Z)}_{\bar{M0^0}}(E_1)P_{\bar{M}^0}(T)LT,
\end{eqnarray}
where $\sigma^{(e^-\,Z)}_{M^0}(E_1)$ and
$\sigma^{(e^+\,Z)}_{\bar{M0^0}}(E_1)$ are the cross sections for the
reactions $e^- + Z \to Z + M^0 + \mu^-$ and $e^+ + Z \to Z + \bar{M}^0
+ \mu^+$, respectively, $E_1$ is the energy of the initial electron
and positron in the laboratory frame, and $L$ is a luminosity of the
Collider.

Calculating the cross sections in the CPT invariant approximation,
$\sigma^{(\vec{e\,}^+\,Z)}_{\bar{M}^0}(E_1) =
\sigma^{(\vec{e\,}^-\,Z)}_{M^0}(E_1)$, the ratio of the numbers of
favourable events $R(T) = N_{M^0}(T)/N_{\bar{M}^0}(T)$ should be
defined only by the ratio $P_{M^0}(T)/P_{\bar{M}^0}(T)$. It reads
\begin{eqnarray}\label{label1.15}
R(T) = \frac{N_{M^0}(T)}{N_{\bar{M}^0}(T)} =
\frac{P_{M^0}(T)}{P_{\bar{M}^0}(T)}.
\end{eqnarray}
Thus, measuring the ratio $R(T)$ of favourable events one can conclude
that (i) CPT invariance is broken if $R(t)$ depends on time of the
observation and oscillates in time, and (ii) CPT invariance is
unbroken if $R(T)$ does not depend on the time of observation. Of
course, this is only a qualitative test.

A practical realization of an experimental test of CPT invariance in
high--energy reactions $e^- + Z \to Z + M^0 + \mu^-$ and $e^+ + Z \to
Z + \bar{M}^0 + \mu^+$ depends on the statistics of favourable events
$N = \sigma L T$ which can be detected during a certain interim of
observation $T$. Nowadays the HERA Collider at DESY operates
$27.5\,{\rm GeV}$ electron and positron beams with luminosities
$L_{e^-} = (15 - 17)\times 10^{30}\,{\rm cm^{-2}\,s^{-1}} = (15 -
17)\,{\rm pb^{-1}}$(H1 $-$ ZEUS) and $L_{e^+} = (65 - 68) \times
10^{30}\,{\rm cm^{-2}\,s^{-1}} = (65 - 68)\,{\rm pb^{-1}}$ (H1 $-$
ZEUS), respectively \cite{HERA}. For these luminosities the number of
events detected during one year for the production of muonium and
anti--muonium are equal to $N_{M^0} = 500\,\sigma_{M^0}$ and
$N_{\bar{M}^0} = 2100\,\sigma_{\bar{M}^0}$, where cross sections
$\sigma_{M^0}$ and $\sigma_{\bar{M}^0}$ are measured in $1\,{\rm pb} =
10^{-36}\,{\rm cm^2}$.

Thus, the problem of an experimental realization of a test of CPT
invariance suggested by Choban and Kazakov \cite{EC01} is related to
(i) the values of the cross sections for the reactions $e^- + Z \to Z
+ X^0 + \mu^-$ and $e^+ + Z \to Z + \bar{X}^0 + \mu^+$, defining total
number of favourable events and (ii) a distinct signal that in the
reactions $e^- + Z \to Z + X^0 + \mu^-$ and $e^+ + Z \to Z + \bar{X}^0
+ \mu^+$ the states $X^0$ and $\bar{X}^0$ should be identified with
muonium $M^0$ and $\bar{M}^0$ anti--muonium, i.e. $X^0 = M^0$ and
$\bar{X}^0 = \bar{M}^0$, respectively.

It is well--known that a more detailed information about nuclear
reactions can be obtained investigating polarizations of coupled
particles. Therefore, in this paper we focus on the calculation of the
cross sections for the high--energy reactions $e^- + Z \to Z + M^0 +
\mu^-$ and $e^+ + Z \to Z + \bar{M}^0 + \mu^+$ in dependence on
polarizations of initial electron and positron and final muons $\mu^-$
and $\mu^+$. Following \cite{MR97} we denote these reactions as
$\vec{e}^{\;-} + Z \to Z + M^0 + \vec{\mu}^{\;-} $ and $\vec{e}^{\;+}
+ Z \to Z + \bar{M}^0 + \vec{\mu}^{\;+}$. We suppose that a dependence
on polarizations of final muons relative to polarizations of initial
electrons and positrons should provide a necessary distinct signal
confirming the production of muonium and anti--muonium with a total
spin $J = 0$ in the reactions $\vec{e}^{\;-} + Z \to Z + X^0 +
\vec{\mu}^{\;-}$ and $\vec{e}^{\;+} + Z \to Z + \bar{X}^0 +
\vec{\mu}^{\;+}$. Indeed, the processes competing with $\vec{e}^{\;-}
+ Z \to Z + X^0 + \vec{\mu}^{\;-} $ and $\vec{e}^{\;+} + Z \to Z +
\bar{X}^0 + \vec{\mu}^{\;+}$ are the reactions $\vec{e}^{\;\mp} + Z
\to Z + \vec{e}^{\;\mp} + \mu^{\;+} + \mu^{\;-}$ of the production of
the $\mu^+\mu^-$ pairs. In these reactions the momenta and
polarizations of $\mu^+$ and $\mu^-$ mesons are strongly correlated
each other and decorrelated with the polarization of the initial
electron (or positron). Therefore, the detection of longitudinally
polarized muons in the final state of the scattering of longitudinally
polarized electrons (or positrons) by a nucleus $Z$ should be a
distinct signal for the production of muonium (or anti--muonium) with
a total spin $J = 0$.

The paper is organized as follows. In section 2 we calculate the
energy spectrum of the final muon and the cross section for the
reaction $\vec{e}^{\;-} + Z \to Z + M^0 + \vec{\mu}^{\;-} $. Since it
is obvious that the CPT violation for the cross sections is negligible
small effect which can be hardly measured, the cross section is
calculated assuming CPT invariance. This implies that the cross
section for the reaction $\vec{e}^{\;-} + Z \to Z + M^0 +
\vec{\mu}^{\;-} $ amounts to the cross section for the reaction
$\vec{e}^{\;+} + Z \to Z + \bar{M}^0 + \vec{\mu}^{\;+}$,
i.e. $\sigma_{M^0}^{(\vec{e}^{\;-} Z)}(E_1) = \sigma_{\bar{M}^0}^{(
\vec{e}^{\;+} Z)}(E_1)$. In Section 3 we estimate the contributions of
the finite nuclear radius and the distortion of the wave functions of
incoming and outcoming leptons caused by the strong Coulomb field
induced by the electric charge $Ze$ with $Z \sim 100$. We estimate
that the contribution of the finite radius of the nucleus is of order
of a few percents. We show that the strong Coulomb field can hardly
destroy the production of bound states of $\mu^+e^-$ and $\mu^-e^+$
pairs, i.e. muoniums and anti--muoniums, in the reactions under
consideration. This is by virtue of the time of the decays $M^0 \to
\mu^+e^-$ and $\bar{M}^0 \to \mu^-e^+$ induced by the strong Coulomb
field is much greater than the time of the production of muonium and
anti--muonium. In the Conclusion we discuss the obtained results and a
practical realization of experiments on the test of CPT invariance for
the HERA Collider at DESY.

\section{Cross sections for reactions $\vec{e}^{\;-} + Z 
\to Z + M^0 + \vec{\mu}^{\;-}$ and $\vec{e}^{\;+} + Z 
\to Z + \bar{M}^0 + \vec{\mu}^{\;+}$} 
\setcounter{equation}{0}

\hspace{0.2in} Feynman diagrams describing the amplitude of the
reaction $\vec{e}^{\;-} + Z \to Z + M^0 + \vec{\mu}^{\;-} $ are
depicted in Fig.1. The amplitude of the reaction $\vec{e}^{\;-} + Z
\to Z + M^0 + \vec{\mu}^{\;-} $ has been calculated in Ref.\cite{EC01}
and reads
\begin{figure}

\begin{center}
 \includegraphics[width=340pt,height=141pt]{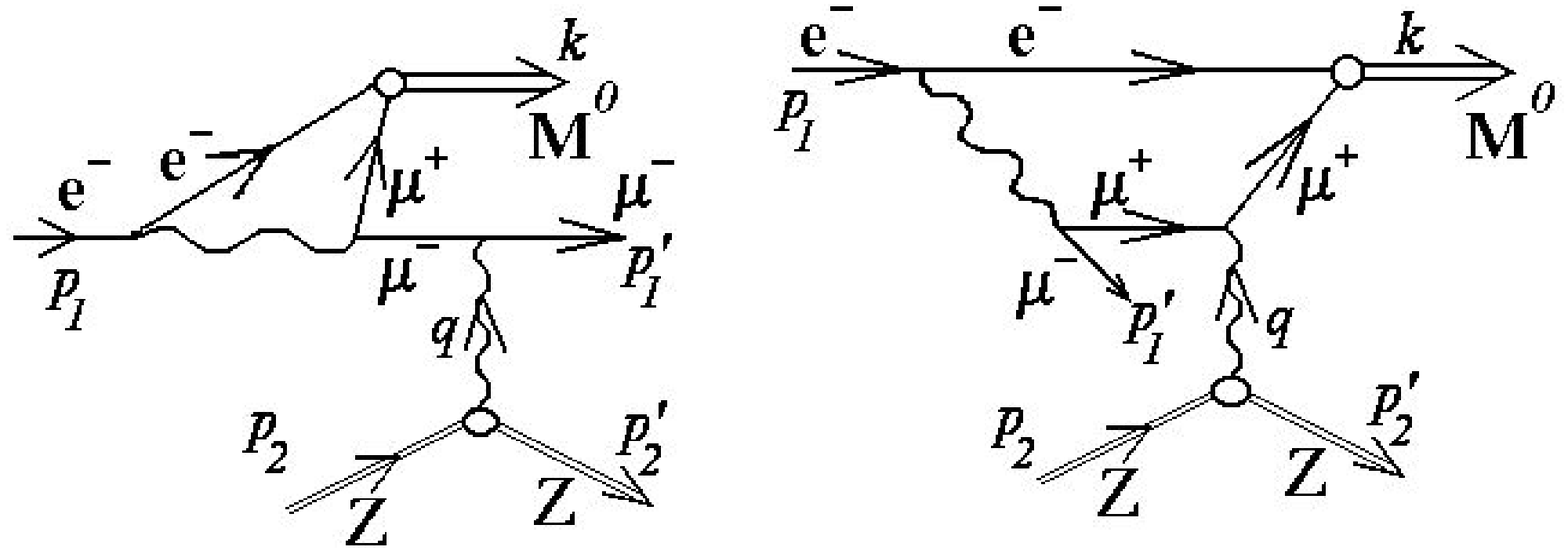}

 Fig. 1.

 Feynman diagrams of the amplitude of the reaction $\vec{e}^{\;-} +
 Z \to Z + M^0 + \vec{\mu}^{\;-} $.
\end{center}

\end{figure}
\begin{eqnarray}\label{label2.1}
\hspace{-0.3in}M(\vec{e}^{\;-}(p_1)Z(p_2) \to Z(p\,'_2)
M^0(k)\vec{\mu}^{\;-}(p\,'_1)) = \frac{\alpha^2}{q^2}\,
\frac{16\pi^2}{m_e}\,\frac{\Psi_{1s}(0)}{m^{3/2}_{\mu}}\,
\frac{{\ell}^{\mu}\, L_{\mu}}{(q^2 - 2q\cdot k)},
\end{eqnarray}
where ${\ell}^{\mu}$ is the electromagnetic current of a nucleus and
$L_{\mu}$ denotes the leptonic current
\begin{eqnarray}\label{label2.2}
L_{\mu} = \bar{u}(p\,'_1,\sigma\,'_1)\,\gamma_5\,(\hat{q}\,p\,'_{1\mu}
- q\cdot p\,'_1\,\gamma_{\mu})u(p_1,\sigma_1),
\end{eqnarray}
where $u(p_1,\sigma_1)$ and $\bar{u}(p\,'_1,\sigma\,'_1)$ are
bispinorial wave functions of an initial electron and final muon
$\mu^-$, $\Psi_{1s}(0) = 1/\sqrt{\pi a^3_B}$ is the wave function of
the muonium in the ground state, $a_B = 1/m_e\alpha = 268.173\,{\rm
MeV}^{-1}$ is the Bohr radius of muonium, $\alpha = 1/137.036$ is the
fine structure constant.

We would like to emphasize that the leptonic current $L_{\mu}$ is
calculated in the ultra--relativistic limit, when masses of leptons
are set zero. According to \cite{EC01} this corresponds to the
kinematical region, where the squared invariant mass of the pair
$M^0\mu^-$, $\omega^2 = (p\,'_1 + k)^2$, is much greater than the
squared mass of the $\mu^-$--meson $m^2_{\mu}$, i.e. $\omega^2 \gg
m^2_{\mu}$. In this kinematical region muonium with a total spin $J =
0$ behaves like a massless neutral scalar point--like particle.

The cross section for the reaction $\vec{e}^{\;-} + Z \to Z + M^0 +
\vec{\mu}^{\;-}$ is defined by
\begin{eqnarray}\label{label2.3}
\hspace{-0.3in}\sigma^{(\vec{e}^{\,-} Z)}_{M^0}(E_1) =
\frac{\alpha^7}{4\pi^2}\frac{m_e}{ m^3_{\mu}}\frac{1}{m_ZE_1}\int
\frac{T_{\mu\nu}R^{\mu\nu}}{q^4(p_1\cdot
p\,'_1)^2}\,\delta^{(4)}(p\,'_2 + p\,'_1 + k - p_2 -
p_1)\,\frac{d^3k}{E}\frac{d^3p\,'_1}{E\,'_1}\frac{d^3p\,'_2}{E\,'_2},
\end{eqnarray}
where $E_1$ is the energy of the initial electron in the laboratory
frame coinciding with the rest frame of a target nucleus $p_{2\mu} =
(m_Z, \vec{0}\,)$, then $E$, $E\,'_1$ and $E\,'_2$ are the energies of
muonium, a $\mu^-$--meson and a final nucleus, respectively. The
tensors $R_{\mu\nu}$ and $T_{\mu\nu}$ are determined by
\begin{eqnarray}\label{label2.4}
\hspace{-0.3in}&&R_{\mu\nu} = \frac{1}{4}\,{\rm Sp}\{(\hat{p}_2 +
m_Z){\ell}^{\dagger}_{\mu}(\hat{p}\,'_2 + m_Z){\ell}_{\nu}\} =
F^2_{1Z}(q^2)\, \Big[2\,p_{2\mu}p_{2\nu} - (p_{2\mu}q_{\nu} +
p_{2\nu}q_{\mu}) + \frac{1}{2}\,q^2\,g_{\mu\nu}\Big]\nonumber\\
\hspace{-0.3in}&&+F^2_{2Z}(q^2)\,\Big[2\,q^2\,m^2_Z\,g_{\mu\nu} +
q^2\,(p_{2\mu}q_{\nu} + p_{2\nu}q_{\mu}) -
q_{\mu}q_{\nu}\,\Big(\frac{1}{2}\,q^2 + 2\,m^2_Z\Big) -
2\,q^2\,p_{2\mu}p_{2\nu}\Big].
\end{eqnarray}
and 
\begin{eqnarray}\label{label2.5}
\hspace{-0.5in}T_{\mu\nu} &=& \frac{1}{4}\,{\rm Sp}\{(\hat{p}_1 -
\gamma_5\hat{w}_1)L^{\dagger}_{\mu}(\hat{p}\,'_1 -
\gamma_5\hat{w}\,'_1)L_{\nu}\} =\nonumber\\
\hspace{-0.5in}&=& \frac{1}{4}\,{\rm Sp}\{(\hat{p}_1 -
\gamma_5\hat{w}_1)\gamma_5(q\cdot p\,'_1\,\gamma_{\mu} -
\hat{q}\,p\,'_{1\mu})(\hat{p}\,'_1 -
\gamma_5\hat{w}\,'_1)\gamma_5(q\cdot p\,'_1\,\gamma_{\nu} -
\hat{q}\,p\,'_{1\nu})\},
\end{eqnarray}
where $F_{1Z}(q^2)$ and $F_{2Z}(q^2)$ are form factors of a nucleus
with a number of protons $Z$.

The polarization matrices $(\hat{p}_1 - \gamma_5\hat{w}_1)$ and
$(\hat{p}\,'_1 - \gamma_5\hat{w}\,'_1)$ are obtained in the zero--mass
limit from the standard polarization matrices $(\hat{p}_1 + m_{\rm
e})(1 - \gamma_5\hat{a})$ and $(\hat{p}\,'_1 + m_{\mu})(1 -
\gamma_5\hat{b})$ \cite{MR97}, where $a_{\mu}$ and $b_{\mu}$,
4--vectors of polarization of the initial electron and the final muon,
are defined by
\begin{eqnarray}\label{label2.6}
a_{\mu} &=& \Big(\frac{\vec{p}_1\cdot \vec{\xi}_1}{m_e}, \vec{\xi}_1
+\frac{\vec{p}_1(\vec{p}_1\cdot \vec{\xi}_1)}{m_e(E_1 +
m_e)}\Big),\nonumber\\ b_{\mu} &=& \Big(\frac{\vec{p}\,'_1\cdot
\vec{\xi}\,'_1}{m_{\mu}}, \vec{\xi}\,'_1
+\frac{\vec{p}\,'_1(\vec{p}\,'_1\cdot \vec{\xi}\,'_1)}{m_{\mu}(E\,'_1
+ m_{\mu})}\Big).
\end{eqnarray}
The 4--vectors of polarization $a_{\mu}$ and $b_{\mu}$ are normalized
by $a_{\mu}a^{\mu} = a^2_0 - \vec{a}^{\,2} = -1$ and $b_{\mu}b^{\mu} =
b^2_0 - \vec{b}^{\,2} = -1$.  In turn, the 3--vectors of polarization
$\vec{\xi}_1$ and $\vec{\xi}\,'_1$ are normalized by
$\vec{\xi}^{\,2}_1 = \vec{\xi}\,'^{\,2}_1 = 1$. Recall that $p_1\cdot
a = p\,'_1\cdot b = 0$.

According to definitions (\ref{label2.6}) the 4--vectors $w_{1\mu}$
and $w\,'_{1\mu}$ are equal to 
\begin{eqnarray}\label{label2.7}
w_{1\mu}&=& (\vec{p}_1\cdot \vec{\xi}_1, \vec{n}_1(\vec{p}_1\cdot
\vec{\xi}_1)) = (\vec{n}_1\cdot \vec{\xi}_1)\,p_{1\mu},\nonumber\\
w\,'_{1\mu} &=& (\vec{p}\,'_1\cdot \vec{\xi}\,'_1,
\vec{n}\,'_1(\vec{p}\,'_1\cdot \vec{\xi}\,'_1)) = (\vec{n}\,'_1\cdot
\vec{\xi}\,'_1)\,p\,'_{1\mu},
\end{eqnarray}
where $\vec{n}_1 = \vec{p}_1/E_1$ and $\vec{n}\,'_1 =
\vec{p}\,'_1/E\,'_1$ and $p_1\cdot w_1 = p\,'_1\cdot w\,'_1 = 0$ due
to $p^2_1 = p\,'\,^2_1 = 0$. The analytical expression of $T_{\mu\nu}$
is given by
\begin{eqnarray}\label{label2.8}
\hspace{-0.3in}T_{\mu\nu}&=&-\,2\,[1 + (\vec{n}_1\cdot
\vec{\xi}_1)(\vec{n}\,'_1\cdot \vec{\xi}\,'_1)]\nonumber\\
\hspace{-0.3in}&&\times\,(p_1\cdot p\,'_1) [(q\cdot
p\,'_1)^2g_{\mu\nu} - (q\cdot p\,'_1)(p\,'_{1\mu}q_{\nu} +
p\,'_{1\nu}q_{\mu}) + q^2p\,'_{1\mu}p\,'_{1\nu}].
\end{eqnarray}
Due to conservation of electric charge the tensors $T_{\mu\nu}$ and
$R_{\mu\nu}$ are gauge invariant
\begin{eqnarray}\label{label2.9}
q^{\mu}T_{\mu\nu} &=& T_{\mu\nu}q^{\nu} = 0,\nonumber\\
q^{\mu}R_{\mu\nu} &=& R_{\mu\nu}q^{\nu} = 0.
\end{eqnarray}
The cross section for the reaction under consideration is then defined
by
\begin{eqnarray}\label{label2.10}
\hspace{-0.3in}&&\sigma^{(\vec{e}^{\,-} Z)}_{M^0}(E_1) =\nonumber\\
\hspace{-0.3in}&&=
\frac{\alpha^7}{\pi^2}\,\frac{m_e}{m^3_{\mu}}\frac{m_Z}{E_1}\int
\frac{(-1)}{q^4(p_1\cdot p\,'_1)}\,[1 + (\vec{n}_1\cdot
\vec{\xi}_1)(\vec{n}\,'_1\cdot \vec{\xi}\,'_1)]\Big\{(F^2_{1Z}(q^2)
-q^2F^2_{2Z}(q^2)) (q\cdot p\,'_1)^2\nonumber\\
\hspace{-0.3in}&& +
\frac{q^2}{m^2_Z}\Big[(F^2_{1Z}(q^2)-q^2F^2_{2Z}(q^2))\Big((p_2\cdot
p\,'_1)^2 - (q\cdot p\,'_1)(p_2\cdot p\,'_1)\Big) +
\frac{1}{2}(F^2_{1Z}(q^2) + 4m^2_ZF^2_{2Z}(q^2))\nonumber\\
\hspace{-0.3in}&& \times\,(q\cdot p\,'_1)^2\Big\}\delta^{(4)}(p\,'_2 +
p\,'_1 + k - p_2 -
p_1)\,\frac{d^3k}{E}\frac{d^3p\,'_1}{E\,'_1}\frac{d^3p\,'_2}{E\,'_2},
\end{eqnarray}
The integration over the phase volume of the final state $ZM^0\mu^-$
we suggest to carry out in the non--relativistic limit of motion of a
final nucleus \cite{AA93}. In this approximation the 4--momentum of
the final nucleus is equal to $p\,'_{2\mu} = (m_Z +
\vec{q}^{\;2}/2m_Z,-\vec{q}\,) = (m_Z + T_2, -\vec{q}\,)$, then the
transferred 4--momentum $q_{\mu} = (- T_2, \vec{q}\,)$ and $q^2 =
-\vec{q}^{\;2}$.

In the non--relativistic limit of motion of a final nucleus the cross
section (\ref{label2.10}) reduces to the form
\begin{eqnarray}\label{label2.11}
\hspace{-0.5in}&&\sigma^{(\vec{e}^{\,-} Z)}_{M^0}(E_1) =
Z^2\,\frac{\alpha^7}{\pi^2}\,\frac{m_e}{m^3_{\mu}}\frac{1}{E_1}\int
\frac{1}{E_1E\,'_1 - \vec{p}_1\cdot\vec{p}\,'_1}\,\Big[1 +
(\vec{n}_1\cdot \vec{\xi}_1)\Big(\frac{\vec{p}\,'_1\cdot
\vec{\xi}\,'_1}{E\,'_1}\Big)\Big]\nonumber\\
\hspace{-0.5in}&& \times\,\Big(E_1{'\,^2} - \frac{(\vec{q}\cdot
\vec{p}\,'_1 )^2}{\vec{q}^{\;2}}\Big)\delta(E\,'_1 + E + T_2 -
E_1)\,\delta^{(3)}(\vec{p}\,'_1 + \vec{k} - \vec{q} -
\vec{p}_1)\,\frac{d^3k}{E}\frac{d^3p\,'_1}{E\,'_1}
\frac{d^3q}{\vec{q}^{\;2}},
\end{eqnarray}
where we have taken into account that $F_{1Z}(0) = Z$ \cite{AA93} and
that the main contribution comes from transferred momenta
$\vec{q}^{\;2}$ comeasurable with zero. The former corresponds to the
Weizs\"acker--Williams approximation \cite{WW34}--\cite{CM94}.

For simplification of the calculation of the phase volume we neglect
the contribution of a kinetic energy of a final nucleus, which is
small compared with typical transferred energies of coupled
leptons. Integrating over $\vec{k}$, a 3--momentum of muonium, we get
\begin{eqnarray}\label{label2.12}
\hspace{-0.5in}&&\sigma^{(\vec{e}^{\,-} Z)}_{M^0}(E_1) =
Z^2\,\frac{\alpha^7}{\pi^2}\,\frac{m_e}{m^3_{\mu}}\frac{1}{E_1}\int
\frac{E\,'_1}{E_1E\,'_1 - \vec{p}_1\cdot\vec{p}\,'_1 }\,\Big[1 +
(\vec{n}_1\cdot \vec{\xi}_1)\Big(\frac{\vec{p}\,'_1\cdot
\vec{\xi}\,'_1}{E\,'_1}\Big)\Big]\nonumber\\
\hspace{-0.5in}&& \times\,\Big(1 - \frac{(\vec{q}\cdot \vec{p}\,'_1
)^2}{\vec{q}^{\;2}E_1{'\,^2}}\Big)\,\delta(E_1 - E\,'_1 -
|\vec{p}\,'_1 - \vec{p}_1 -\vec{q}\,|
)\,\frac{d^3p\,'_1}{|\vec{p}\,'_1 - \vec{p}_1
-\vec{q}\,|}\,\frac{d^3q}{\vec{q}^{\;2}}=\nonumber\\ 
\hspace{-0.5in}&& =
Z^2\,\frac{\alpha^7}{\pi^2}\,\frac{m_e}{m^3_{\mu}}\frac{1}{E_1}\int
\Big[1 + (\vec{n}_1\cdot \vec{\xi}_1)\Big(\frac{\vec{p}\,'_1\cdot
\vec{\xi}\,'_1}{E\,'_1}\Big)\Big]\,\frac{E\,'_1\,
I(\vec{p}_1,\vec{p}\,'_1)}{E_1E\,'_1 - \vec{p}_1\cdot\vec{p}\,'_1
}\,d^3p\,'_1,
\end{eqnarray}
where we have denoted
\begin{eqnarray}\label{label2.13}
I(\vec{p}_1,\vec{p}\,'_1) = \int\Big(1 - \frac{(\vec{q}\cdot
\vec{p}\,'_1 )^2}{\vec{q}^{\;2}E_1{'\,^2}}\Big)\,\delta(E_1 - E\,'_1 -
|\vec{p}\,'_1 - \vec{p}_1 -\vec{q}\,| )\,\frac{1}{|\vec{p}\,'_1 -
\vec{p}_1 -\vec{q}\,|}\,\frac{d^3q}{\vec{q}^{\;2}}.
\end{eqnarray}
The integration over $\vec{q}$ we carry out assuming that
$|\vec{p}\,'_1 - \vec{p}_1| \gg |\vec{q}\,|$ that is valid for the
Weizs\"acker--Williams approximation. Using a vector $\vec{z} =
\vec{q}/|\vec{p}\,'_1 - \vec{p}_1|$ we obtain
\begin{eqnarray}\label{label2.14}
I(\vec{p}_1,\vec{p}\,'_1) = \int\Big(1 - \frac{(\vec{z}\cdot
\vec{p}\,'_1 )^2}{\vec{z}^{\;2}E_1{'\,^2}}\Big)\,\delta\Big(E_1 -
E\,'_1 - |\vec{p}\,'_1 - \vec{p}_1| +(\vec{p}\,'_1 - \vec{p}_1)\cdot
\vec{z}\, \Big)\,\frac{d^3z}{\vec{z}^{\;2}}.
\end{eqnarray}
Now it is convenient to introduce new variables $x = E\,'_1/E_1$,
$\vec{n}\,'_1 = \vec{p}\,'_1/E\,'_1$ and $\vec{n}_1 =
\vec{p}_1/E_1$. In these variables the function
$I(\vec{p}_1,\vec{p}\,'_1)$ reads
\begin{eqnarray}\label{label2.15}
I(\vec{p}_1,\vec{p}\,'_1) = \frac{1}{E_1}\int\Big(1 -
\frac{(\vec{z}\cdot \vec{n}\,'_1
)^2}{\vec{z}^{\;2}}\Big)\,\delta\Big(1 - x - |x\vec{n}\,'_1 -
\vec{n}_1| +(x\vec{n}\,'_1 - \vec{n}_1)\cdot \vec{z}\,
\Big)\,\frac{d^3z}{\vec{z}^{\;2}}.
\end{eqnarray}
The next step in the integration over $\vec{z}$ is to rewrite the
integral in the following form
\begin{eqnarray}\label{label2.16}
\hspace{-0.7in}&&I(\vec{p}_1,\vec{p}\,'_1) = \nonumber\\
\hspace{-0.7in}&&=\frac{1}{\pi E_1}{{\cal
R}e}\int\limits^{\infty}_0d\lambda\,e^{\textstyle i\lambda(1-x -
|x\vec{n}\,'_1 - \vec{n}_1|)}\int\Big(1 +
\frac{1}{\lambda^2\vec{z}^{\;2}}\frac{\partial^2}{\partial
x^2}\Big)\,e^{\textstyle i\lambda(x\vec{n}\,'_1 - \vec{n}_1)\cdot
\vec{z}}\,\frac{d^3z}{\vec{z}^{\;2}}.
\end{eqnarray}
Since the integrals over $\vec{z}$ are equal to
\begin{eqnarray}\label{label2.17}
\hspace{-0.3in}&&\int e^{\textstyle i\lambda(x\vec{n}\,'_1 -
\vec{n}_1)\cdot \vec{z}}\;\frac{d^3z}{\vec{z}^{\;2}} =
\frac{4\pi}{\lambda|x\vec{n}\,'_1 -
\vec{n}_1|}\int\limits^{\infty}_0\frac{\sin z}{z}\,dz
=\frac{4\pi}{\lambda|x\vec{n}\,'_1 - \vec{n}_1|}\lim_{\alpha \to
1}\int\limits^{\infty}_0\frac{\sin z}{z^{\alpha}}\,dz =\nonumber\\
\hspace{-0.3in}&&= \frac{4\pi}{\lambda|x\vec{n}\,'_1 -
\vec{n}_1|}\lim_{\alpha \to 1}{{\cal
I}m}\int\limits^{\infty}_0dz\,e^{\textstyle i z}z^{-\alpha} =
\frac{4\pi}{\lambda|x\vec{n}\,'_1 - \vec{n}_1|}\lim_{\alpha \to
1}{{\cal I}m}\,\frac{\Gamma(1-\alpha)}{(-i)^{1-\alpha}} =\nonumber\\
\hspace{-0.3in}&&= \frac{4\pi}{\lambda|x\vec{n}\,'_1 -
\vec{n}_1|}\,\lim_{\alpha \to
1}\Gamma(2-\alpha)\,\frac{\displaystyle\sin\Big(\frac{\pi}{2}
(1-\alpha)\Big)}{1-\alpha} = \frac{2\pi^2}{\lambda|x\vec{n}\,'_1 -
\vec{n}_1|}
\end{eqnarray}
and 
\begin{eqnarray}\label{label2.18}
\hspace{-0.3in}&&\int e^{\textstyle i \lambda(x\vec{n}\,'_1 -
\vec{n}_1)\cdot\vec{z} }\;\frac{d^3z}{\vec{z}^{\;4}} = \nonumber\\
\hspace{-0.3in}&&= 4\pi \lambda|x\vec{n}\,'_1 - \vec{n}_1|
\int\limits^{\infty}_0\frac{\sin z}{z^3}dz = 4\pi
\lambda|x\vec{n}\,'_1 - \vec{n}_1|\lim_{\alpha \to 3}{{\cal
I}m}\int\limits^{\infty}_0dz\,e^{\textstyle i z}z^{-\alpha}
=\nonumber\\ \hspace{-0.3in}&&= 4\pi \lambda|x\vec{n}\,'_1 -
\vec{n}_1|\lim_{\alpha \to 3}{{\cal
I}m}\,\frac{\Gamma(1-\alpha)}{(-i)^{1-\alpha}} = 4\pi
\lambda|x\vec{n}\,'_1 - \vec{n}_1|\lim_{\alpha \to
3}\Gamma(1-\alpha)\sin\Big(\frac{\pi}{2}(1-\alpha)\Big) =\nonumber\\
\hspace{-0.3in}&&= 4\pi \lambda|x\vec{n}\,'_1 - \vec{n}_1|\lim_{\alpha
\to 3}\Gamma(4-\alpha)\frac{\displaystyle
\sin\Big(\frac{\pi}{2}(1-\alpha)\Big)}{(1-\alpha)(2-\alpha)(3 -
\alpha)}=\nonumber\\
\hspace{-0.3in}&&= - 4\pi \lambda|x\vec{n}\,'_1 -
\vec{n}_1|\lim_{\alpha \to 3}\Gamma(4-\alpha)\frac{\displaystyle
\sin\Big(\frac{\pi}{2}(3-\alpha)\Big)}{(1-\alpha)(2-\alpha)(3 -
\alpha)} = - \pi^2 \lambda|x\vec{n}\,'_1 - \vec{n}_1|,
\end{eqnarray}
the function $I(\vec{p}_1,\vec{p}\,'_1)$ is defined by the integral
over $\lambda$
\begin{eqnarray}\label{label2.19}
I(\vec{p}_1,\vec{p}\,'_1) =
\frac{\pi}{E_1}\Big(\frac{1}{|x\vec{n}\,'_1 - \vec{n}_1|} + \frac{(x
-\vec{n}\,'_1\cdot \vec{n}_1)^2 }{|x\vec{n}\,'_1 -
\vec{n}_1|^3}\Big)\int\limits^{\infty}_0\frac{d\lambda}{\lambda}\,
\cos(\lambda(1-x - |x\vec{n}\,'_1 - \vec{n}_1|)).
\end{eqnarray}
The integral over $\lambda$ is divergent. However, it can be
regularized by following the theory of generalized functions
\cite{GS64}. The result reads
\begin{eqnarray}\label{label2.20}
I(\vec{p}_1,\vec{p}\,'_1) =
\frac{\pi}{E_1}\Big(\frac{1}{|x\vec{n}\,'_1 - \vec{n}_1|} + \frac{(x
-\vec{n}\,'_1\cdot \vec{n}_1)^2 }{|x\vec{n}\,'_1 -
\vec{n}_1|^3}\Big){\ell n}\Big(\frac{1}{|x\vec{n}\,'_1 - \vec{n}_1| -
(1 -x)}\Big).
\end{eqnarray}
Substituting (\ref{label2.20}) in (\ref{label2.12}) and proceeding to
variables $x$ and $\vec{n}\,'_1$ we define the energy spectrum of a
final $\mu^-$--meson
\begin{eqnarray}\label{label2.21}
\hspace{-0.5in}\frac{1}{x^2}\frac{d\sigma^{(\vec{e}^{\,-}
Z)}_{M^0}(E_1)}{dx} &=&
Z^2\,\frac{\alpha^7}{\pi}\,\frac{m_e}{m^3_{\mu}}\int \frac{1 +
(\vec{n}_1\cdot \vec{\xi}_1)(\vec{n}\,'_1\cdot \vec{\xi}\,'_1)}{1 -
\vec{n}_1\cdot\vec{n}\,'_1 }\,\Big(\frac{1}{|x\vec{n}\,'_1 -
\vec{n}_1|} + \frac{(x -\vec{n}\,'_1\cdot \vec{n}_1)^2
}{|x\vec{n}\,'_1 - \vec{n}_1|^3}\Big)\nonumber\\ &&\times\,{\ell
n}\Big(\frac{1}{ |x\vec{n}\,'_1 - \vec{n}_1|-
(1-x)}\Big)\,d\Omega_{\,\vec{n}\,'_1}.
\end{eqnarray}
For the subsequent integration over a unit vector $\vec{n}\,'_1$ we
introduce angular variables as follows
\begin{eqnarray}\label{label2.22}
\vec{n}_1\cdot\vec{n}\,'_1 &=& \cos\vartheta_1\,',\nonumber\\
\vec{n}\,'_1\cdot \vec{\xi}\,'_1&=& \cos\vartheta_1\,'\cos\Theta_1\,'
+\sin\vartheta_1\,'\sin\Theta_1\,'\cos(\varphi_1\,' -
\Phi_1\,'\,),\nonumber\\ d\Omega_{\,\vec{n}\,'_1} &=&
\sin\vartheta_1\,'d\vartheta_1\,'d\varphi_1\,',
\end{eqnarray}
where $\Theta_1\,'$ and $\Phi_1\,'$ are polar and azimuthal angles of
the polarization vector $\vec{\xi}\,'_1$ relative to the momentum
$\vec{p}_1$. In (\ref{label2.22}) we have taken into account that
$|\vec{\xi}\,'_1| = 1$. Integrating over $\varphi_1\,'$ we get
\begin{eqnarray}\label{label2.23}
\hspace{-0.5in}&&\frac{1}{x^2}\frac{d\sigma^{(\vec{e}^{\,-}
Z)}_{M^0}(E_1)}{dx} = 2 Z^2 \alpha^7
\frac{m_e}{m^3_{\mu}}\int\limits^{\pi}_0 \frac{1 + (\vec{n}_1\cdot
\vec{\xi}_1)\cos\vartheta_1\,'\cos\Theta_1\,'}{1 -
\cos\vartheta_1\,'}\,\Big(\frac{1}{\sqrt{1 - 2x\cos\vartheta_1\,' +
x^2}}\nonumber\\
\hspace{-0.5in}&& + \frac{(x -\cos\vartheta_1\,')^2 }{(1 -
2x\cos\vartheta_1\,' + x^2)^{3/2}}\Big)\,{\ell n}\Big(\frac{1}{\sqrt{1
- 2x\cos\vartheta_1\,' + x^2} -(1-x)}\Big)\,
\sin\vartheta_1\,'d\vartheta_1\,'
\end{eqnarray}
Now it is convenient to introduce a new variable $t = \sqrt{1 -
2x\cos\vartheta_1\,' + x^2}$, which varies in the limits $1-x \le t \le
1+x$. In terms of $t$ the energy spectrum (\ref{label2.23}) reads
\begin{eqnarray}\label{label2.24}
\hspace{-0.5in}\frac{1}{x}\frac{d\sigma^{(\vec{e}^{\,-}
Z)}_{M^0}(E_1)}{dx} &=& 2 Z^2 \alpha^7 \frac{m_e}{m^3_{\mu}}
\int\limits^{1+x}_{1-x} \frac{2x + (1+ x^2 - t^2)(\vec{n}_1\cdot
\vec{\xi}_1)\cos\Theta_1\,'}{t^2 - (1-x)^2}\nonumber\\
&&\times\,\Big(1+ \frac{(1 - x^2 - t^2)^2}{4x^2t^2}\Big)\,{\ell
n}\Big(\frac{1}{t - (1-x)}\Big)\,dt.
\end{eqnarray}
It is seen that the integral over $t$ is concentrated in the vicinity
of the lower limit. The singularity of the integrand in the vicinity
of the lower limit can be easily regularized by making a change of the
lower limit $1 - x \to 1 - x + \Lambda^2/E^2_1$, where $\Lambda$ is a
cut--off restricting energies of a final $\mu^-$--meson from
below. According to the kinematical region $\omega^2 \gg m^2_{\mu}$
\cite{EC01} the cut--off $\Lambda$ can be chosen of order of $\Lambda
\simeq 1\,{\rm GeV}$. Such a dependence on the cut--off $\Lambda$ can
be justified as follows: $E\,'_1 = |\vec{p}\,'_1| =
\sqrt{(\vec{p}\,'_1)^2 + \Lambda^2 - \Lambda^2} =
\sqrt{(\vec{p}\,'_1)^2 + \Lambda^2} - \Lambda^2/\sqrt{(\vec{p}\,'_1)^2
+ \Lambda^2} \to E\,'_1 - \Lambda^2/E\,'_1 \approx E\,'_1 -
\Lambda^2/E_1$.

Keeping only the dominant contributions to the integral over $t$ we
get
\begin{eqnarray}\label{label2.25}
\frac{d\sigma^{(\vec{e}^{\,-} Z)}_{M^0}(E_1)}{dx} = 8 Z^2 \alpha^7
\frac{m_e}{m^3_{\mu}}\,{\ell
n}^2\Big(\frac{E_1}{\Lambda}\Big)\,\frac{x^2}{1-x}\,[1 +
(\vec{n}_1\cdot \vec{\xi}_1)\cos\Theta_1\,'\,].
\end{eqnarray}
Introducing the angle $\Theta_1$, defined by $\vec{n}_1\cdot
\vec{\xi}_1 = \cos\Theta_1$, where we have taken into account that
$|\vec{\xi}_1| =1$, we obtain the energy spectrum of $\mu^-$--mesons
for the reaction $\vec{e}^{\;-} + Z \to Z + M^0 + \vec{\mu}^{\;-}
$ in dependence on the polarizations of the initial electron and the
final $\mu^-$--muon described by the angles $\Theta_1$ and
$\Theta\,'_1$
\begin{eqnarray}\label{label2.26}
\frac{d\sigma^{(\vec{e}^{\,-} Z)}_{M^0}(E_1)}{dx} = 8 Z^2 \alpha^7
\frac{m_e}{m^3_{\mu}}\,{\ell
n}^2\Big(\frac{E_1}{\Lambda}\Big)\,\frac{x^2}{1-x}\,(1 +
\cos\Theta_1\cos\Theta_1\,'\,).
\end{eqnarray}
Integrating over $x$ we arrive at the total cross section for the
reaction $\vec{e}^{\;-} + Z \to Z + M^0 + \vec{\mu}^{\;-} $
\begin{eqnarray}\label{label2.27}
\sigma^{(\vec{e}^{\,-} Z)}_{M^0}(E_1) = 16 Z^2 \alpha^7
\frac{m_e}{m^3_{\mu}}\,{\ell n}^3\Big(\frac{E_1}{\Lambda}\Big)\,(1 +
\cos\Theta_1\cos\Theta_1\,'\,).
\end{eqnarray}
Assuming that electrons are longitudinally polarized electrons,
$\cos\Theta_1 = 1$, one can see that for the fixed electron energy the
cross section acquires the maximal value only for longitudinally
polarized muons $\cos\Theta_1\,' = 1$. This agrees with the production
of muonium with a total spin $J = 0$. Thus, we argue that the
appearance of longitudinally polarized muons in the final state of the
reaction $\vec{e}^{\;-} + Z \to Z + X + \vec{\mu}^{\;-} $ should
testify the production of muonium $X \equiv M^0$.

For the numerical estimate of the cross sections at the energies
available for the HERA Collider at DESY \cite{HERA}, i.e $E_1 =
27.5\,{\rm GeV}$, we suggest to use Radon, ${^{222}_{~86}}{\rm Rn}$,
as a target nucleus, since Radon has a spin 1/2. The cross sections
for longitudinally polarized electrons and positrons scattering by
${^{222}_{~86}}{\rm Rn}$ and longitudinally polarized muons are equal
to
\begin{eqnarray}\label{label2.28}
\sigma^{(\vec{e}^{\;-}{\rm Rn})}_{M^0}(E_1 = 27.5\,{\rm GeV}) =
\sigma^{(\vec{e}^{\;+}{\rm Rn})}_{\bar{M}^0}(E_1 = 27.5\,{\rm GeV}) =
1.6\,{\rm pb}.
\end{eqnarray}
In our calculation the cross section for the reaction $e^- + Z \to Z +
M^0 + \mu^-$ has turned out to be dependent on a cut--off $\Lambda
\simeq 1\,{\rm GeV}$. In this connection we would like to remind that
the problem of the appearance of a cut--off in cross sections for some
reactions calculated within the Weizs\"acker--Williams approximation
has been pointed out by Bertulani and Baur \cite{CB88}.

Now let us discuss the energy dependence of the cross section
(\ref{label2.27}). It is well--known that for the $e^+e^-$ pair
production in heavy--ion collisions \cite{CB88}--\cite{JE90} and
$p\bar{p}$ collisions \cite{CM94} the cross section for a capture of a
final electron in an atomic $K$--shell orbit is proportional to ${\ell
n}(\gamma_{\rm coll})$, where $\gamma_{\rm coll}$ is a Lorentz factor
of colliding particles in the center of mass frame. This factor is
related to the corresponding Lorentz factor $\gamma_p$ of the
projectile (for a fixed target machine) by $\gamma_p = 2\gamma^2_{\rm
coll} - 1$ \cite{CB88,JE90}, where $\gamma_p \sim E_1$.

In turn, the cross section for the production of a point--like neutral
scalar particle in high--energy heavy--ion collisions in the
Weizs\"acker--Williams approximation is proportional to ${\ell
n}^3(\gamma_{\rm coll})$ \cite{CB88,JE90}.

For very high energies, when masses of coupled leptons can be
neglected, muonium with a total spin $J = 0$ can be treated as a
point--like massless scalar neutral particle. Such a property of
muonium is caused by an addition pole--singularity appearing at $(q -
k)^2 = q^2 - 2k\cdot q = 0$ for $k^2 = m^2_{\mu} = 0$ (see
Eq.(2.1)). This makes the part of the diagram in Fig.1, responsible
for creation of muonium, equivalent to an amplitude of a process
$\gamma^* + \gamma^* \to M^0$, where $\gamma^*$'s are virtual
photons. That is why the obtained cross section for the reaction $e^-
+ Z \to Z + M^0 + \mu^-$ has turned out to be proportional to ${\ell
n}^3(\gamma_{\rm coll})$.

\section{Influence of  a finite radius of a nucleus and 
a distortion of wave functions of coupled leptons}
\setcounter{equation}{0}

In this Section we estimate the influence of a finite radius of the
nucleus $Z$. According to \cite{AA93} the form factor of the nucleus
with a mass number $A$ can be defined by the expansion
\begin{eqnarray}\label{label3.1}
\frac{1}{Z}\,F_{1Z}(q^2) = 1 - \frac{1}{6}\,r^2_A\,\vec{q}^{\;2} +
O(\vec{q}^{\;4})
\end{eqnarray}
where we identify $r_A$ with a radius of a nucleus with a mass number
$A$ given by \cite{AA93}
\begin{eqnarray}\label{label3.2}
r_{\rm A} = 1.2\,A^{1/3}\,{\rm fm} = 6.1\,A^{1/3}\,{\rm GeV}^{-1}.
\end{eqnarray}
Due to the finite value of the nuclear radius the function
$I(\vec{p}_1,\vec{p}\,'_1)$ changes as follows
\begin{eqnarray}\label{label3.3}
\delta I(\vec{p}_1,\vec{p}\,'_1) &=& -
\frac{\pi}{E_1}\,\frac{1}{3}\,r^2_{\rm A} E^2_1\Big(-
\frac{1}{|x\vec{n}\,'_1 - \vec{n}_1|} + 3\,\frac{(x -\vec{n}\,'_1\cdot
\vec{n}_1)^2 }{|x\vec{n}\,'_1 - \vec{n}_1|^3}\Big)(1 - x -
|x\vec{n}\,'_1 - \vec{n}_1|)^2\nonumber\\ &&\times {\ell
n}\Big(\frac{1}{|x\vec{n}\,'_1 - \vec{n}_1| - (1 - x)}\Big).
\end{eqnarray}
In the region of the integration over $t$, dominant for the leading
term of the expansion of the form factor $F_{1Z}(\vec{q}^{\,2})$ into
the powers of $\vec{q}^{\,2}$, the contribution of the finite radius
of the nucleus can be summarized as
\begin{eqnarray}\label{label3.4}
\sigma^{(\vec{e}^{\,-} Z)}_{M^0}(E_1) = \frac{16 Z^2
\alpha^7}{\displaystyle \Big(1 + \frac{1}{6}\frac{r^2_{\rm
A}\Lambda^4}{E^2_1}\Big)^2}\,\frac{m_e}{m^3_{\mu}}\,{\ell
n}^3\Big(\frac{E_1}{\Lambda}\Big)\,(1 +
\cos\Theta_1\cos\Theta_1\,'\,).
\end{eqnarray}
For the electron (positron) scattering by ${^{222}_{~86}}{\rm Rn}$
with the laboratory energy $E_1 = 27.5\,{\rm GeV}$ the correction to
the cross section, caused by the finite value of the nucleus radius
(\ref{label3.2}), can be made of order of 4$\%$ varying the parameter
$\Lambda$ from $\Lambda \simeq 1\,{\rm GeV}$ to $\Lambda \simeq
0.8\,{\rm GeV}$ in comparison with the value of the cross section
(\ref{label2.28}) calculated for $E_1 = 27.5\,{\rm GeV}$, $\Lambda
\simeq 1\,{\rm GeV}$ and $r_{\rm A} = 0$. Hence, in the
Weizs\"acker--Williams approximation \cite{WW34}--\cite{CM94}
without loss of generality we can treat a nucleus $Z$ in the reactions
$\vec{e}^{\,-} + Z \to Z + M^0 + \vec{\mu}^{\,-}$ and $\vec{e}^{\,+} +
Z \to Z + \bar{M}^0 + \vec{\mu}^{\,+}$ as a point--like particle with
the electric charge $Ze$.

In the strong Coulomb field caused by a point--like charge $Ze$ for
$Z\sim 100$ the wave functions of the initial electron (positron) and
the final muon should be distorted. According to \cite{AA93} at very
high energies and in the eiconal approximation these wave functions
can be written in the following form
\begin{eqnarray}\label{label3.5}
\hspace{-0.3in}\Psi_{e^{\,-}}(\vec{r}_1;\vec{p}_1, \sigma_1)_{\,\rm
in} &=& u(\vec{p}_1, \sigma_1)\exp\Big\{+ i\vec{p}_1\cdot \vec{r}_1 +
i\,\frac{E_1}{|\vec{p}_1|}\int^{\infty}_0
\frac{Ze^2\,ds}{\sqrt{\vec{\rho}^{\;2}_1 + (z -
s)^2}}\Big\},\nonumber\\
\hspace{-0.3in}\Psi_{\mu^{\,-}}(\vec{r}\,'_1; \vec{p}\,'_1,
\sigma\,'_1)_{\,\rm out} &=& u(\vec{p}\,'_1, \sigma\,'_1)\exp\Big\{+
i\vec{p}\,'_1\cdot \vec{r}\,'_1 -
i\,\frac{E\,'_1}{|\vec{p}\,'_1|}\int^{\infty}_0
\frac{Ze^2\,ds}{\sqrt{\vec{\rho}^{\;'2}_1 + (z\,' + s)^2}}\Big\},
\end{eqnarray}
where $\vec{\rho}_1$ and $\vec{\rho}\,'_1$ are components of
radius--vectors $\vec{r}_1$ and $\vec{r}\,'_1$ perpendicular to the
momentum $\vec{p}_1$ and $\vec{p}\,'_1$, respectively.

In the limit $m_e = m_{\mu} = 0$ the wave functions (\ref{label3.5})
change themselves as
\begin{eqnarray}\label{label3.6}
\hspace{-0.3in}\Psi_{e^{\,-}}(\vec{r}; \vec{p}_1, \sigma_1)_{\,\rm in}
&=& u(\vec{p}_1, \sigma_1)\exp\Big\{+ i\vec{p}_1\cdot \vec{r} +
i\int^{\infty}_0 \frac{Ze^2\,ds}{\sqrt{\vec{\rho}^{\;2} + (z -
s)^2}}\Big\},\nonumber\\
\hspace{-0.3in}\Psi_{\mu^{\,-}}(\vec{r}; \vec{p}\,'_1,
\sigma\,'_1)_{\,\rm out} &=& u(\vec{p}\,'_1, \sigma\,'_1)\exp\Big\{+
i\vec{p}\,'_1\cdot \vec{r} - i\int^{\infty}_0
\frac{Ze^2\,ds}{\sqrt{\vec{\rho}^{\;2} + (z + s)^2}}\Big\},
\end{eqnarray}
where we have taken into account the fact that at high energies
effectively the production of the final muon occurs at the same
spatial point $\vec{r}_1 = \vec{r}\,'_1 = \vec{r}$, where the initial
electron has been absorbed. The amplitude of the reaction
$\vec{e}^{\;-} + Z \to Z + M^0 + \vec{\mu}^{\;-} $ is proportional to
the product
\begin{eqnarray}\label{label3.7}
\hspace{-0.3in}&&\Psi^{\dagger}_{\mu^{\,-}}(\vec{r}; \vec{p}\,'_1,
\sigma\,'_1)_{\rm in}\Psi_{e^{\,-}}(\vec{r}; \vec{p}_1, \sigma_1)_{\rm
out}\sim \nonumber\\
\hspace{-0.3in}&& \sim \exp\Big\{i\int^{\infty}_0
\frac{Ze^2\,ds}{\sqrt{\vec{\rho}^{\;2} + (z + s)^2}} +
i\int^{\infty}_0 \frac{Ze^2\,ds}{\sqrt{\vec{\rho}^{\;2} + (z -
s)^2}}\Big\} = \nonumber\\
\hspace{-0.3in}&&= \exp\Big\{i\int^{\infty}_{-\infty}
\frac{Ze^2\,ds}{\sqrt{\vec{\rho}^{\;2} + s^2}}\Big\} = e^{\textstyle
-i\,Ze^2\,{\ell n}[C\vec{\rho}^{\;2}]},
\end{eqnarray}
where $C$ is an undefined constant related to the large--distance
regularization of the integrals in (\ref{label3.7}). The spinorial
factor has been taken already into account for the calculation of the
cross section (\ref{label2.27}) or (\ref{label3.4}).

Formally the amplitude of the reaction $\vec{e}^{\;-} + Z \to Z + M^0
+ \vec{\mu}^{\;-}$ in the momentum representation should be obtained
by means of the integration over configuration space that includes the
integration over $\vec{\rho}$ as well. However, due to the presence of
the undefined infinitesimal constant $C$, the integration over
$\vec{\rho}$ can be reduced to the replacement of $\vec{\rho}^{\;2}$
by an average value.

Since $|\vec{\rho}\,|$ is a transversal scale of the reaction
$\vec{e}^{\;-} + Z \to Z + M^0 + \vec{\mu}^{\;-}$, which can be
treated as an impact parameter of this reaction, for an estimate of an
average value of this parameter we can set $\vec{\rho}^{\;2} \sim
\sigma^{(\vec{e}^{\;-}Z)}(E_1)_{\rm max} \sim {\ell
n}^3(E_1/\Lambda)$. This yields
\begin{eqnarray}\label{label3.8}
\Psi^{\dagger}_{\mu^{\,-}}(\vec{r}; \vec{p}\,'_1,
\sigma\,'_1)\Psi_{e^{\,-}}(\vec{r}; \vec{p}_1, \sigma_1) \sim
e^{\textstyle -i\,Ze^2\,{\ell n}[C\,'{\ell n}^3(E_1/\Lambda)]}.
\end{eqnarray}
As the cross section for the reaction is proportional to
$|\Psi^{\dagger}_{\mu^{\,-}}(\vec{r}; \vec{p}\,'_1,
\sigma\,'_1)\Psi_{e^{\,-}}(\vec{r}; \vec{p}_1, \sigma_1)|^2$, the
distortion of the wave functions of the initial and final leptons
caused by the strong Coulomb field does not change crucially the cross
section for the reaction $\vec{e}^{\;-} + Z \to Z + M^0 +
\vec{\mu}^{\;-}$ calculated for the wave functions of the coupled
leptons in the form of plane waves.

For the estimate of the influence of the strong Coulomb field on the
state of muonium we suggest to calculate the time of the decay $M^0
\to \mu^+ + e^-$ induced by the external Coulomb field. The amplitude
of the decay $M^0 \to \mu^+ + e^-$ we define as
\begin{eqnarray}\label{label3.9}
{\cal M}(M^0 \to \mu^+ + e^-) = \int \frac{1}{\sqrt{V}}\,e^{\textstyle
-i\vec{p}\cdot \vec{r}}\,U(\vec{r}\,)\,\frac{1}{\sqrt{\pi
a^3_{B}}}\,e^{\textstyle - r/a_{B}}\, d^3r,
\end{eqnarray}
where $\vec{p}$ is a relative momentum of the $\mu^+ e^-$ pair, $a_{B}
= 268.173\,{\rm MeV}^{-1}$ is the Bohr radius of muonium, and $V$ is a
normalization volume. Then, $U(\vec{r}\,)$ is the potential energy of
the dipole moment $\vec{d} = e\,\vec{r}$ of the $\mu^+ e^-$ pair
coupled to the strong Coulomb field of the nucleus $Z$
\begin{eqnarray}\label{label3.10}
U(\vec{r}\,) = - \vec{d}\cdot \vec{E}(\vec{r}\,) = \frac{Ze^2}{r}.
\end{eqnarray}
Integrating over $\vec{r}$ we get
\begin{eqnarray}\label{label3.11}
{\cal M}(M^0 \to \mu^+ + e^-) = \frac{4\pi Z e^2}{\sqrt{V\pi
a^3}}\,\frac{a^2_{B}}{1 + a^2_{B}p^2}.
\end{eqnarray}
The time of the decay $M^0 \to \mu^+ + e^-$ is equal to
\begin{eqnarray}\label{label3.12}
\tau^{-1}(M^0 \to \mu^+e^-) = \frac{32Z^2\alpha^2}{a^3_{B} E^2_{M^0}}
= \frac{32Z^2\alpha^5}{E^2_{M^0}}\Big(\frac{m_e m_{\mu}}{m_e +
m_{\mu}}\Big)^3,
\end{eqnarray}
where $E_{M^0}$ is the energy of the muonium in the rest frame of the
nucleus $Z$. Since $E_{M^0} \gg 5\,{\rm GeV}$, for ${^{222}_{~86}}{\rm
Rn}$ we estimate $\tau(M^0 \to \mu^+e^-) \gg 2.6 \times 10^{-8}\,{\rm
s}$. The time of the interaction of the electron scattering by Radon,
during which muonium can be produced, is of order $\tau \sim
10^{-8}\,{\rm s}$. This means that the strong Coulomb field does not
affect crucially the production of muonium or anti--muonium in the
reactions $\vec{e}^{\;-} + Z \to Z + M^0 + \vec{\mu}^{\;-}$ and
$\vec{e}^{\;+} + Z \to Z + \bar{M}^0 + \vec{\mu}^{\;+}$. Of course, a
more detailed analysis of the Coulomb distortion of the wave functions
of leptons in the reactions $\vec{e}^{\;-} + Z \to Z + M^0 +
\vec{\mu}^{\;-} $ and $\vec{e}^{\; +} + Z \to Z + \bar{M}^0 +
\vec{\mu}^{\;+}$ and the influence of this distortion on the
production of muonium $M^0$ and anti--muonium $\bar{M}^0$ is required.
We are planning to analyse this problem in our forthcoming
investigations.

\section{Conclusion}
\setcounter{equation}{0}

\hspace{0.2in} We have calculated the cross sections for the reactions
$\vec{e}^{\;-} + Z \to Z + M^0 + \vec{\mu}^{\;-} $ and $\vec{e}^{\; +}
+ Z \to Z + \bar{M}^0 + \vec{\mu}^{\;+}$ of the production of muonium
$M^0$ and anti--muonium $\bar{M}^0$ with polarized $\mu^-$ and $\mu^+$
mesons by polarized electrons and positrons coupled at high energies
to the nucleus $Z$.

The cross sections are calculated in dependence on (i) an energy $E_1$
of initial electron and positron in the laboratory frame, coinciding
with the rest frame of a target nucleus $Z$, and (ii) polarizations of
initial electron and positron and final muons in the kinematical
region $\omega^2 = (p_1\,' + k)^2 \gg m^2_{\mu}$ making the massless
limit of coupled leptons reasonable.

For the numerical estimate of the cross sections at the energies
available for the HERA Collider at DESY \cite{HERA}, i.e $E_1 =
27.5\,{\rm GeV}$, we suggest to use Radon, ${^{222}_{~86}}{\rm Rn}$,
as a target nucleus, since Radon has a spin 1/2. The theoretical
values of the cross sections for longitudinally polarized electrons
and positrons scattering by ${^{222}_{~86}}{\rm Rn}$ are equal to
$\sigma^{(\vec{e}^{\;-}{\rm Rn})}_{M^0}(E_1 = 27.5\,{\rm GeV}) =
\sigma^{(\vec{e}^{\;+}{\rm Rn})}_{\bar{M}^0}(E_1 = 27.5\,{\rm GeV}) =
1.6\,{\rm pb}$. For these cross sections we predict the following
numbers of favourable events: $N_{M^0} = 808$ and $N_{\bar{M}^0} =
3360$. Hence, the increase of luminosities of electron and positron
beams should make the experiment for a test of CPT invariance,
suggested by Choban and Kazakov in Ref.\cite{EC01}, feasible at DESY.

We have estimated the influence of the finite nuclear radius and the
Coulomb distortion of the wave functions of the leptons. According to
our estimate in the kinematical region $\omega^2 = (p_1\,' + k)^2 \gg
m^2_{\mu}$ the Weizs\"acker--Williams approach, treating a nucleus as
a point--like particle and neglecting the Coulomb distortion of the
wave functions of incoming and outcoming leptons, is a rather
well--defined approximation. The contribution of the finite nuclear
radius can be kept at the level of a few percents. The distortion of
the wave functions of the initial and final leptons caused by the
strong Coulomb field does not change the cross sections for the
reactions under consideration. During the time of the production of
muonium or anti--muonium the strong Coulomb field induced by the
charge of the nucleus $Ze$ does not destroy bound states of $\mu^+e^-$
or $\mu^-e^+$ pairs. Hence, the strong Coulomb field can hardly screen
the phenomenon of the violation of CPT invariance in the reactions
$\vec{e}^{\;-} + Z \to Z + M^0 + \vec{\mu}^{\;-}$ and $\vec{e}^{\;+} +
Z \to Z + \bar{M}^0 + \vec{\mu}^{\;+}$.

We have shown that the test of CPT invariance in the reactions
$\vec{e}^{\;-} + Z \to Z + M^0 + \vec{\mu}^{\;-}$ and $\vec{e}^{\;+} +
Z \to Z + \bar{M}^0 + \vec{\mu}^{\;+}$ reduces to the experimental
analysis of the ratio $R(T) = N_{M^0}(T)/N_{\bar{M}^0}(T)$
(\ref{label1.15}) of the numbers of favourable events detected during
an interim $T$. If $R(T)$ is a constant in time -- CPT invariance is
conserved, and if $R(T)$ is an oscillating function in time one can
conclude that CPT invariance is violated. 

We would like to accentuate that this is a qualitative analysis of CPT
invariance. In the case of the ratio $R(T)$ oscillating in time we can
infer neither a strength nor a nature of a violation of CPT
invariance.

We argue that the appearance of longitudinally polarized muons in the
final states of the reactions $\vec{e}^{\;-} + Z \to Z + X +
\vec{\mu}^{\;-}$ and $\vec{e}^{\;+} + Z \to Z + \bar{X} +
\vec{\mu}^{\;+}$ with longitudinally polarized electrons and positrons
is a distinct signal for the production of muonium $M^0$ and
anti--muonium $\bar{M}^0$ with a total spin $J = 0$. This should
testify that $X \equiv M^0$ and $ \bar{X} \equiv \bar{M}^0$ with a
total spin $J = 0$.

Indeed, the creation of the $\mu^+\mu^-$ pairs in the reactions
$\vec{e}^{\;\mp} + Z \to Z + \vec{e}^{\;\mp} + \mu^+ + \mu^-$ seems to
be the main process competing with the production of muonium and
anti--muonium in the reactions $\vec{e}^{\;-} + Z \to Z + M^0 +
\vec{\mu}^{\;-}$ and $\vec{e}^{\;+} + Z \to Z + \bar{M}^0 +
\vec{\mu}^{\;+}$. The main distinction of the production of the
$\mu^+\mu^-$ pairs from the production of muonium and anti--muonium is
a strong correlation between the momenta and polarizations of $\mu^+$
and $\mu^-$ and a decorrelation of them with the initial electron or
positron. In turn, a strong correlation between the polarizations of
the final muons and the initial electron and positron is a feature of
the production of muonium and anti--muonium with a total spin $J = 0$
in the reactions $\vec{e}^{\;-} + Z \to Z + M^0 + \vec{\mu}^{\;-}$ and
$\vec{e}^{\;+} + Z \to Z + \bar{M}^0 + \vec{\mu}^{\;+}$. Hence, at
first glimpse for the experimental realization of the test of CPT
invariance in the reactions $\vec{e}^{\;-} + Z \to Z + M^0 +
\vec{\mu}^{\;-}$ and $\vec{e}^{\;+} + Z \to Z + \bar{M}^0 +
\vec{\mu}^{\;+}$ with longitudinally polarized electrons and positrons
it suffices to count the number of longitudinally polarized $\mu^-$
and $\mu^+$ mesons during an interim $T$. Plotting the ratio of these
numbers, which should coincide with $R(T)$, one should obtain an
experimental information about CPT invariance.


\newpage

\begin{thebibliography}{9}
\bibitem{KH02} 
K. Hagiwara {\it et al.}, 
Phys. Rev. D {\bf 66}, 010001 (2002).
\bibitem{SW95} 
S. Weinberg, in {\it THE QUANTUM THEORY OF FIELDS}, {\it
Foundations} Vol. I, Cambridge University Press, 1995; {\it Modern
Applications} Vol. II, Cambridge University Press, 1996;
{\it Supersymmetry} III, Cambridge University Press, 2000.
\bibitem{AW80} 
R. F. Streater and A. S. Wightman,
in {\it PCT, SPIN AND STATISTICS, AND ALL THAT},
Third Edition, Princeton University Press, Princeton and Oxford, 
1980.
\bibitem{PDG1}
(see \cite{KH02} p.313)
\bibitem{LS69}
L. M. Sehgal,
Phys. Rev. {\bf 181}, 2151 (1969);
J. P. Hsu, 
Phys. Rev. D {\bf 5}, 981 (1972); Phys. Rev. D {\bf 9}, 304 (1974);
R. Morse, U. Nauenberg, E. Bierman, D. Sager, and A. P. Colleraine,
Phys. Rev. Lett. {\bf 28}, 388 (1972);
S. Barshay,
Phys. Lett. B {\bf 101}, 155 (1981);
W. Bernreuther, U. Law, J. P. Ma, and O. Nachtmann,
Z. Phys. C {\bf 41}, 143 (1988);
G. Gabrielse,
Nucl. Phys. Proc. Suppl. {\bf 8}, 448 (1989);
{\it Parity and Time Reversal Violation in Compound Nuclear 
States and Related Topics}, edited by A. Auerbach and J. D. 
Bowman, World Scientific, Singapore, 1996;
P. Colangelo and  G. Corcela,
Eur. Phys. J. C {\bf 1}, 515 (1998);
S. R. Coleman and  S. L. Glashow Phys. Rev. 
D {\bf 59}, 116008 (1999).
\bibitem{MG86}
{\it UNIFIED STRING THEORIES},
edited by M. Green and D. Gross,
World Scientific, Singapore, 1986;
{\it SUPERSTINGS, A Theory of Everything}?,
edited by P. C. W. Davies and J. Brown, Cambridge University Press, 
1988;
B. F. Hatfield,
in {\it QUANTUM FIELD THEORY OF POINT PARTICLES AND STRINGS},
 Frontiers in Physics, Addison--Wesley Publishing Co., Singapore,
1989;
B. M. Barbashov and V. V. Nesterenko,
in {\it INTRODUCTION TO THE RELATIVISTIC STRING THEORY},
World Scientific, Singapore, 1990;
L. Castellani, R. D'Auria, and P. Fr$\acute{\rm e}$,
in {\it SUPERGRAVITY AND SUPERSTRINGS, A Geometric Perspective}, 
{\it Superstrings}, Vol. 3, World Scientific, Singapore, 1991;
J. Polchinski,
in {\it STRING THEORY, Superstring Theory and Beyond}, Vol. II,
Cambridge University Press, 1998.
\bibitem{VK91}
V. A. Kosteleck$\acute{\rm y}$ and R. Potting,
Nucl Phys. B {\bf 359}, 545 (1991); Phys. Lett. B {\bf 381},
(1996); D. Colladay and V. A. Kosteleck$\acute{\rm y}$, 
Phys. Rev. D {\bf 55}, 6760 (1997), Phys. Rev. D {\bf 58}, 
116002 (1998); {\it CPT AND LORENTZ SYMMETRY}, edited by 
V. A. Kosteleck$\acute{\rm y}$, World Scientific, Singapore, 1999;
V. A. Kosteleck$\acute{\rm y}$ and C. D. Lane,
Phys. Rev. D {\bf 60}, 116010 (1999).
\bibitem{GL77}
G. P. Lepage,
Phys. Rev. A {\bf 16}, 863 (1977).
\bibitem{VH92}
V. W. Hughes,
Z. Phys. {\bf 56}, S35 (1992).
\bibitem{CPT1}
D. Kawall, V. W. Hughes, M. G. Perdekamp, W. Liu, K. P. Jungmann,
and G. Zu Putlitz,
{\it Test of CPT and Lorentz Invariance from Muonium Spectroscopy},
hep--ex/0201010.
\bibitem{CPT2}
L. Willmann and K. P. Jungmann,
{\it The Muonium Atom as a Probe of Physics beyond the 
Standard Model}, Physics {\bf 499}, 43 (1997), hep-ex/9805013;
K. P. Jungmann,
{\it Searching New Physics in Muonium Atoms}, hep--ex/9805015.
\bibitem{EC01}
G. A. Kazakov and E. A. Choban,
JETP Letters {\bf 74}, 216 (2001).
\bibitem{SS61}
S. S. Schweber,
in {\it AN INTRODUCTION TO RELATIVISTIC QUANTUM FIELD THEORY},
Row, Peterson and Co$\,\bullet\,$ Evanston, Ill.,
Elmsford, New York, 1961.
\bibitem{IZ80}
C. Itzykson and J.--B. Zuber,
in {\it QUANTUM FIELD THEORY}, McGraw--Hill Book Co.,
New York, 1980, p.154.
\bibitem{HERA}
Roberto Sacchi,
{\it Search for Physics Beyond the Standard Model at HERA}, 
DESY, 2002.
\bibitem{MR97}
 M. P. Rekalo, J. Arvieux, and E. Tomasi--Gustafsson,
Phys. Rev. C {\bf 56}, 2238 (1997);
 A. Ya. Berdnikov, Ya. A. Berdnikov, A. N. Ivanov, V. A. Ivanova,
V. F. Kosmach, M. D. Scadron, and N. I. Troitskaya, Eur. Phys. J. A
{\bf 12} (2000) 341, hep--ph/0110050.
\bibitem{AA93}
A. I. Akhiezer, A. G. Sitenko, and V. K. Tartakovskii,
in {\it NUCLEAR ELECTRODYNAMICS}, Springer--Verlag, Berlin, 1993.
\bibitem{WW34}
C. F. von Weizs\"acker, Z. Phys. {\bf 88}, 612 (1934);
E. J. Williams, Phys. Rev. {\bf 45}, 729 (1934).
\bibitem{VG61} 
V. N. Gribov, V. A. Kolkunov,  L. B. Okun, and V. M. Shekhter, 
JETP {\bf 41}, 1839 (1961).
\bibitem{CB88} 
C. A. Bertulani and G. Baur,
Phys. Rep. {\bf 163}, 299 (1988).
\bibitem{RA87}
R. Anholt and U. Becker,
Phys. Rev. A {\bf 36}, 4628 (1987).
\bibitem{JE90}
J. Eichler,
Phys. Rep. {\bf 193}, 165 (1990).
\bibitem{CM94}
C. T. Munger, S. J. Brodsky, and I. Schmidt,
Phys. Rev. D {\bf 49}, 3228 (1994).
\bibitem{GS64} 
I. M. Gel'fand and G. E. Shilov,
in {\it GENERALIZED FUNCTIONS, Properties and Operations}, Vol. I, 
Academic Press, New York and London, 1964.
\end{thebibliography}
\end{document}